\documentclass[aps,pra,twocolumn,twoside,a4paper,amsfonts,showpacs]{revtex4-1}
\usepackage{amsmath,graphicx,epsfig,dsfont}
\usepackage[english]{babel}
\renewcommand{\.}{\hspace{-0.5ex}}
\renewcommand{\vec}[1]{\mathbf{#1}}
\renewcommand{\S}[1]{\mathcal{S}}
\newcommand{\E}[1]{\mathcal{E}}
\newcommand{\SE}[1]{\mathcal{SE}}
\newcommand{\bra}[1]{\langle#1|}
\newcommand{\ket}[1]{|#1\rangle}

\newcommand{\expect}[1]{\langle#1 \rangle}

\newcommand{\tr}{\text{Tr}}
\renewcommand{\Re}{\text{Re}}

\newcommand{\kom}[2]{\left[#1,#2\right]}

\newcommand{\tot}{\text{tot}}
\newcommand{\PT}{\text{PT}}
\newcommand{\n}{\overline{n}}

\newcommand{\sys}{\text{sys}}
\newcommand{\env}{\text{env}}
\newcommand{\hc}{\text{hc.}}
\newcommand{\red}{\text{red}}

\newcommand{\inter}{\text{int}}

\newcommand{\ent}{\text{ent}}

\newcommand{\crit}{\text{crit}}
\newcommand{\dec}{\text{dec}}
\newcommand{\therm}{\text{therm}}

\begin{document}
\title{Decoherence and the Nature of System-Environment Correlations}
\author{A.~Pernice}
\email{Ansgar.Pernice@tu-dresden.de}
\author{W.T.~Strunz}
\affiliation{Institut f\"ur Theoretische Physik, Technische Universit\"at Dresden, D-01062 Dresden, Germany}

%
\begin{abstract}
We investigate system-environment correlations based on the exact dynamics 
of a qubit and its environment in the framework of pure decoherence
(phase damping).
We focus on the relation of decoherence and the build-up of system-reservoir 
entanglement for an arbitrary (possibly mixed) initial qubit state.
In the commonly employed regime where the qubit dynamics can be 
described by a Markov
master equation of Lindblad type, we find that for almost all qubit initial states inside the
Bloch sphere, decoherence
is complete while the total state is still separable - no entanglement is 
involved.
In general, both ``separable'' and ``entangling'' decoherence occurs, 
depending on temperature and initial qubit state.
Moreover, we find situations where classical and quantum correlations 
periodically
alternate as a function of time in the regime of low temperatures.
\end{abstract}
\date{\today}

\pacs{03.65.Yz,42.50.Lc,03.65.Ta}

\maketitle

\section{Introduction}
The increasing possibilities and sophisticated ways to control larger and larger quantum
systems require a thorough understanding
of the dynamics of open quantum systems~\cite{weiss-2008,breuer+petruccione-2002}.
As prime example, the phenomenon of decoherence is an effect of the 
environmental impact on quantum behavior of the system, leading 
to a -- sometimes rapid -- decay of its coherences in a certain
basis~\cite{joos+zeh-2003,zurek-2003,strunz-2002,schlosshauer-2008,braun-2001}.

With respect to quantum information processing, decoherence is the main
obstacle that needs to be overcome. However,
the influence of the environment on an open quantum system is not always of 
disturbing nature. For example, 
an environment can also mediate quantum correlations between two
qubits~\cite{braun-2002,yuan+kuang+liao-2010}, which implies that in the 
process system and environment must get quantum mechanically correlated. 
It is an interesting question under which 
circumstances and how exactly those correlations develop.
Many interesting results about correlations and entanglement in quantum many 
body systems have been obtained in the last 
few years~\cite{eisert+cramer+plenio-2010}.
The issue of system-environment correlations is also of importance for 
the quantum foundations of
thermodynamics~\cite{mahler-2007,hilt+lutz-2009,popescu+short+winter-2006}.

Although there are methods to detect the formation of system-environment correlations at the level of reduced 
dynamics~\cite{kimura+al-2007}, a detailed analysis of the role of \emph{quantum} correlations in decoherence phenomena lies 
beyond such an approach.
To gain deeper understanding one surely has to investigate the total 
(sometimes thermal) state 
of the composite quantum system, which in fact
has been done for a couple of model 
systems~\cite{maziero+al-2010,lopez+al-2008,hilt+lutz-2009,mcaneney+al-2003,eisert+plenio-2002,khemmani+sa-yakanit+strunz-2005}. Some of 
the authors find decoherence accompanied by system-reservoir entanglement, some of them report of certain conditions under 
which decoherence appears without. Still, we lack a complete picture
of entanglement {\it dynamics} including finite temperature environments and 
arbitrary system initial states.

Here we present an approach to this problem, based on a partial 
P-representation of the total state, investigating a purely 
decohering qubit. It enables us to make detailed, state and 
temperature dependent statements on the evolution of the correlations between 
system and environment, which we relate to the decoherence process.

\section{Open Quantum System Model for Decoherence and Reduced Dynamics}
Describing pure decoherence (phase damping), we will assume a standard 
model~\cite{feynman+vernon-1963,caldeira+leggett-1981} $H_\tot=H_\sys+H_\inter+H_\env$,
by coupling a qubit $H_\sys=\frac{\hbar \Omega}{2}\sigma_z$ via
$H_\inter=\sigma_z\otimes(\sum_{\lambda=1}^N g_\lambda^*a_\lambda^\dagger+\hc)$
non-dissipatively to a bath of harmonic oscillators
$H_\env=\sum_{\lambda=1}^N\hbar\omega_\lambda a_\lambda^\dagger
a_\lambda$~\cite{skinner+hsu-1986,unruh-1995,kuang+zeng+tong-1999,yu+eberly-2002,yuan+kuang+liao-2010}.
Here $\Omega$ denotes the transition frequency of the qubit and the coefficients $g_\lambda$ describe the coupling
strength between the qubit and each environmental mode of frequency
$\omega_\lambda$.
Assuming the environment to be initially in the thermal state
$\rho_\therm=\exp[-H_\env/k_B T]/\tr[\exp[-H_\env/k_B T]]$
at temperature $T$, and the total initial state to be the product $\rho_\tot(0)=\rho_{\sys}(0)\otimes\rho_\therm$, the following exact time local master equation for the reduced density operator $\rho_\red=\tr_\env[\rho_\tot]$
 follows
\begin{equation}
\dot{\rho}_\red=-i\frac{\Omega}{2}\kom{\sigma_z}{\rho_\red}-\frac{\gamma(t)}{2}\left(\rho_\red-\sigma_z\,\rho_\red\,\sigma_z\right).
 \label{equ:masterequation-decoherence}
\end{equation}
Its solution reads
\begin{equation}
\rho_\red(t)=\begin{pmatrix}
\rho_{00}&\mathcal{D}(t)\rho_{01}\\
\mathcal{D}^*(t)\rho_{10}&\rho_{11}
\end{pmatrix},
\label{equ:solution-roh_red}
\end{equation}
with $\mathcal{D}(t)=\exp[-i\Omega t-\int_0^t\gamma(s)ds]$. Here $\rho_{ij}$
represents the {\it initial} state of the qubit,
with Bloch vector $\vec{r}=(x,y,z)=\expect{\vec{\sigma}}$.
For the following, it is essential to consider \emph{arbitrary} (mixed) initial qubit states. 
This choice reflects experimental limitations in the preparation process, and also allows us to investigate
decoherence and qubit-environment correlations in cases where the qubit is part of an entangled many qubit
state~\cite{yu+eberly-2003,mintert+al-2005}.
While pure initial qubit states will always lead to system-environment entanglement, even a slight
decrease of initial purity may have a significant effect on the nature of the ensuing system-environment correlations.

In~(\ref{equ:masterequation-decoherence}) and (\ref{equ:solution-roh_red}) we have introduced the time dependent decoherence rate $\gamma(t)$ which 
by means of the spectral density of the environment
$J(\omega)=\sum_{\lambda=0}^N|g_\lambda|^2\delta(\omega-\omega_\lambda)$
can be written as
\begin{equation}
\gamma(t)=4\int_0^t\.ds\int_0^\infty \.\.d\omega
J(\omega)\coth[\hbar\omega/2 k_B T]\cos[\omega s].
\label{equ:decoherence_rate}
\end{equation}
For a time independent $\gamma(t)\equiv\gamma$, qubit coherences decay on the decoherence time scale $\tau_\dec=\gamma^{-1}$, sometimes referred to as ``$T_2$''.
For the general case considered here, we define the decoherence time $\tau_\dec$ through
the relation $\int_0^{\tau_\dec}\gamma(s)ds=1$. The off-diagonal elements of the qubit have then decayed to a fraction
$e^{-1}$ of the initially present quantum coherences.

Indeed, a constant decoherence rate $\gamma(t)\equiv\gamma$ is obtained for 
an Ohmic spectral density
with a cut-off frequency $\omega_c$, in the limit of high
temperatures $k_B T\gg\hbar\omega_c$ and large times $\omega_c t\gg 1$.
Then eq.~(\ref{equ:masterequation-decoherence}) is
of Lindblad type, and the corresponding dynamics is Markovian. 
For our studies we choose a sharp cut-off, $J(\omega)=\kappa\,\omega\,\Theta(\omega-\omega_c)$~\cite{caldeira+leggett-1983}, where $\kappa$ parametrises the coupling strength between system and environment.
Accordingly, the asymptotic decoherence rate is given by $\hbar\gamma=4\pi\kappa\,k_BT$.

Decoherence of qubits is sometimes modelled by random unitary dynamics~\cite{yu+eberly-2003,helm+strunz-2010}.
It is known that on the level of the reduced state, single-qubit
decoherence can always be modelled that way, while for two qubits or more,
there is genuine ``quantum'' decoherence -- i.e. there is decoherence 
that -- even on the reduced level --
can only 
be modelled by coupling the system to a quantum environment
~\cite{landau+streater-1993,buscemi+chiribella+dariano-2005,helm+strunz-2009}.
Considering a single qubit and its given quantum
environment, this
raises the question to what extent the total state involves any 
quantum correlations.

In fact, the most general class of total initial states which allows for 
completely positive
reduced dynamics is classically correlated, with zero discord~\cite{rodriguez-rosario+al-2008,shabani+lidar-2009}.
We emphasize that for our special choice of the spectral density, the map $\rho_\red(t_i)\rightarrow\rho_\red(t_f)$ induced by~(\ref{equ:masterequation-decoherence}) is completely positive for all $0<t_i<t_f$.
This is due to the positivity of
$\gamma(t)$ in our case, which also implies the "Markovianity" of the
reduced dynamics for all times in the sense of both works~\cite{breuer+laine+piilo-2009,rivas+huelga+plenio-2010}.

For all these recent investigations it is relevant to gain a deeper understanding
of the build-up of quantum correlations between system and environment.
In this paper we investigate the correlation dynamics of the total state
for a mixed initial product state and identify entangled
and classically correlated regimes.
It turns out that in general the dynamics of system-environment correlations can
be markedly rich.
Let us provide the framework necessary to extract the desired information from
the total state.

\section{The Total State}
We are interested 
in an exact expression for the total state $\rho_\tot(t)$, which we will represent in a coherent state
basis. In a sense, our approach here is a mixed-state generalization of the \emph{non-Markovian
quantum state diffusion} approach to open quantum system dynamics~\cite{diosi+strunz-1997,strunz-2005}.
We will combine the environmental coherent state labels
into the vector $\xi=(\xi_1,\xi_2,\cdots)$ of complex numbers and consistently make use of
the notation $d^2\xi/\pi:=d^2\xi_1/\pi\; d^2\xi_2/\pi\cdots$ (see 
also~\cite{strunz-2005}). Furthermore, introducing the average thermal occupation
number $\n_\lambda$ of the $\lambda$-th 
environmental mode, it is possible to expand the total state in terms of coherent states 
$\ket{\xi}$
\begin{equation}
 \rho_\tot(t)=\int\frac{d^2\xi}{\pi}\frac{1}{\n}e^{-|\xi|^2/\n}\;
\hat{P}(t;\xi,\xi^*)\otimes\ket{\xi}\bra{\xi},
\label{equ:P-representation}
\end{equation}
defining the matrix-valued partial P-function $\hat{P}(t)$ with values in the
$2\times2$ dimensional state space of the qubit. Here we symbolically write
$\exp[-|\xi|^2/\n]/\n:=\prod_\lambda\exp[-|\xi_\lambda|^2/\n_\lambda]/\n_\lambda$. 

We find that expression~(\ref{equ:P-representation}) is a solution of
the total von-Neumann equation
with initial $\rho_\tot(0)=\rho_\sys(0)\otimes\rho_\therm$ if 
\begin{equation}
\hat{P}(t;\xi,\xi^*)=
\begin{pmatrix}
\mathcal{A}^+(t;\xi,\xi^*)\rho_{00}&\mathcal{B}(t;\xi,\xi^*)\rho_{01}\\
\mathcal{B}^*(t;\xi,\xi^*)\rho_{10}&\mathcal{A}^-(t;\xi,\xi^*)\rho_{11}
\end{pmatrix},
\label{equ:exact-solution}
\end{equation}
where
$\mathcal{A}^{\pm}=\exp[-A(t)\pm\left\{(a(t)|\xi)+(\xi|a(t))\right\}]$
and 
$\mathcal{B}=\exp[-i\Omega
t]\exp[B(t)-\left\{({b}(t)|\xi)-(\xi|b(t))\right\}]$.
Here we have introduced the complex time dependent vectors
${a}(t)=(a_1(t),a_2(t),\cdots)$ and ${b}(t)$ with scalar product $(a(t)|\xi)\equiv\sum_\lambda a_\lambda^*(t)\xi_\lambda$
and vector components
\begin{eqnarray*}
 a_\lambda(t)&=&\frac{1}{\n_\lambda}\int_0^t\left(g_\lambda e^{i\omega_\lambda s}\right)ds\\
 b_\lambda(t)&=&\frac{2\n_\lambda+1}{\n_\lambda}\int_0^t\left(g_\lambda e^{i\omega_\lambda s}\right)ds.
\end{eqnarray*}
Furthermore, we use the abbreviations
\begin{eqnarray*}
A(t)&=&2\,\Re\.\int_0^tds\int_0^sd\tau\left[\sum_\lambda\frac{1}{\n_\lambda}|g_\lambda|^2e^{-i\omega_\lambda(t-s)}\right]\\
B(t)&=&2\,\Re\.\int_0^tds\int_0^sd\tau\left[\sum_\lambda\frac{2\n_\lambda+1}{\n_\lambda}|g_\lambda|^2e^{-i\omega_\lambda(t-s)}\right].
\end{eqnarray*}
Initially, $\hat P = \rho_\sys(0) = \rho_\red(0)$ and
note that there are no approximations necessary to achieve the
result~(\ref{equ:exact-solution}) and thus via~(\ref{equ:P-representation}) to 
obtain the exact state of the composite system.
\vspace{2em}

\section{System-Environment Separability and Decoherence}
\begin{figure}
\vspace{4ex}
\includegraphics[width=0.47\textwidth]{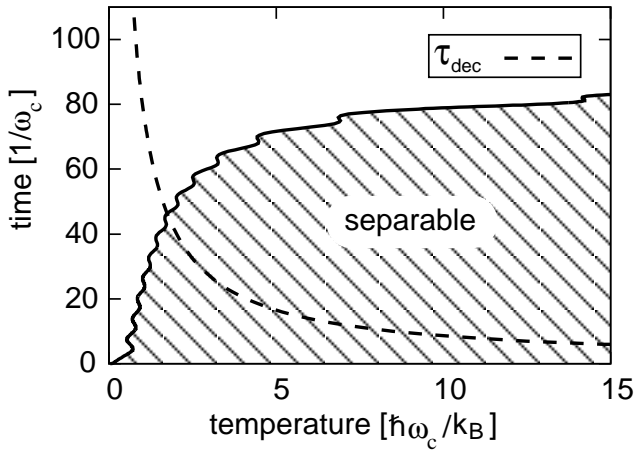}
 \caption{Separability of system and
environment vs decoherence time - Lindblad regime: 
for temperatures $k_B T\gg\hbar\omega_c$ the qubit may decohere without becoming
entangled to its environment despite an initial purity of $r = 0.98$.}
 \label{fig:sep-ent-dec_hightemp}
\end{figure}

Having the total state at hand we now want to draw conclusions on its
separability. For this purpose we argue as follows:

As long as the partial P-function is positive-semi-definite the total state
$\rho_\tot(t)$ in representation~(\ref{equ:P-representation}) is trivially separable. Since one eigenvalue of $\hat P$ is
always positive, a good indicator for this circumstance is the determinant
$\det[\hat{P}]$ which in fact is independent of the coherent state labels $\xi$ and $\xi^*$:
\begin{equation}
 \det[\hat{P}(t)]=e^{-2A(t)}\left[1-z^2\right]-e^{2B(t)}(x^2+y^2).
\label{equ:positive-P}
\end{equation}
Recall that $(x,y,z)$ are the Bloch coordinates of the \emph{initial} state of the qubit.
This expression is positive as long as the following condition is fulfilled
\begin{equation}
 S(t):=A(t)+B(t)\le\frac{1}{2}\ln\left[\frac{1-z^2}{x^2+y^2}\right].
\label{equ:condition-sep}
\end{equation}
The quantity on the l.h.s. depends on the temperature of the surrounding heat
bath via the thermal occupation numbers $\n_\lambda=(\exp[\hbar\omega/k_BT]-1)^{-1}$,
and reads
\begin{equation}
\begin{split}
 S(T,t)=4&\int_0^t\.ds\int_0^s\.d\tau\int_0^\infty\.\.d\omega\\
 &\quad\times J(\omega)\;\exp\left[\hbar\omega/kT\right]\cos[\omega(s-\tau)].
\end{split}
\label{equ:S_t_continuum}
\end{equation}
With our special choice of $J(\omega)$, $S(T,t)$ can be written in terms of known special functions.
For a given initial qubit state with $\expect{\vec\sigma}=(x,y,z)$, condition~(\ref{equ:condition-sep})
defines an area in
the $(T,t)$-plane, shown in fig.~\ref{fig:sep-ent-dec_hightemp}, within which the total state is necessarily separable.
Note that for a pure initial state the r.h.s. of~(\ref{equ:condition-sep}) is zero and the separable area vanishes.

It turns out that as $t\rightarrow\infty$, $S(T,t)$ grows above all bounds for all temperatures and it becomes
independent of temperature in the high temperature
limit, $S(T,t)\rightarrow S(t)\approx\ln\,t+\text{const}$. As a consequence, for any initial state and temperature,
expression~(\ref{equ:P-representation}) of the total state at some point in
time ceases to represent a proper mixture and thus, separability can no longer be concluded.
Let us stress that positivity of $\hat P(t)$ is a sufficient but not a 
necessary condition for separability.

\subsection*{Separability in the Lindblad Regime}
Condition~(\ref{equ:condition-sep}) is visualized in fig.~\ref{fig:sep-ent-dec_hightemp}, where we relate our
criterion for separability to the decoherence time $\tau_\dec$ introduced earlier.
We resort to the \emph{weak coupling} regime
($\kappa=10^{-3}$) and choose a qubit initial state in the
equatorial plane ($z=0$) of the Bloch sphere with purity $r=0.98$ 
(recall that $r$ denotes the length
of the Bloch vector, i.e., $r = |\vec r|$).

While at low temperatures $\tau_\dec$ lies outside of
our separability area, it is inside this area for temperatures of
about $k_B T\ge 2\,\hbar\omega_c$. In the high temperature regime, when the dynamics is governed by a Lindblad
equation, decoherence of the qubit is thus
complete, while the total state is still separable. This is
especially striking because the initial state is so close to being the
pure state $(|0\rangle+ |1\rangle)/\sqrt{2}$. 

\begin{figure}
\vspace{4ex}
\includegraphics[width=0.47\textwidth]{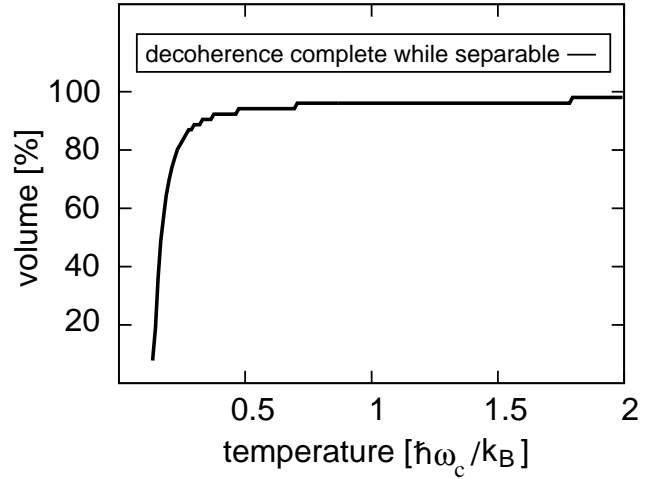}
 \caption{Fraction of qubit initial states that decohere while the total state is still separable:
remarkably, in the regime of validity of the Lindblad equation
($k_B T\gg\hbar\omega_c$) virtually all initial states
decohere before any entanglement to the bath is built up.}
\label{fig:volumeofstates}
\end{figure}
The question arises as to how many initial states, depending on
temperature, meet the very same fate.
In fig.~\ref{fig:volumeofstates} we display the relative volume of the set of states that
are still separable at $\tau_\dec$, i.e. after decoherence.
Remarkably, in the regime where the master equation~(\ref{equ:masterequation-decoherence})
is of Lindblad form ($\gamma=\text{const}$) for practically all states decoherence has long
happened before any entanglement between system an environment can build up.

In fact, this is not too surprising: in the weak coupling limit 
(``Born approximation'') a master equation 
of Lindblad type is usually derived assuming that the joint state of system plus
environment to the lowest order in the interaction remains a product
state~\cite{breuer+petruccione-2002,pechukas-1994}:
one substitutes $\rho_\tot(t)\approx\rho_\sys(t)\otimes\rho_\therm$ in the
perturbative evolution equation. The
severity of this approximation is apparently manifested in the
fact that entanglement builds up very slowly compared to the time scale given by decoherence.

\section{System-Environment Entanglement and Decoherence}
All pure qubit initial states different from $\ket{0}$ and $\ket{1}$ will lead to system-environment entanglement
immediately. However, as we have shown in the previous section even a slight reduction of purity (e.g. $r=0.98$) may
lead to a prolonged period of separability, long enough for the qubit to loose its coherence.
Being separable at least up to the time when $\det[\hat{P}]$ becomes negative, $\rho_\tot$
turns into an entangled state at the latest when its partial transpose $\rho_\tot^\PT$
yields a negative expectation value $\mathcal{E}^\PT=\bra{\Psi}\rho_\tot^\PT\ket{\Psi}$ in some state
$\ket{\Psi}$ of the composite system~\cite{peres-1996}. Hence, to detect
entanglement we aim to find a state $\ket{\psi(t;\xi^*)}\sim\expect{\xi\,|\Psi(t)}$
in the Hilbert space of the qubit such that
\begin{equation}
 \mathcal{E}^\PT\sim\int\frac{d^2\xi}{\pi}\frac{1}{\n}e^{-|\xi|^2/\n}\bra{\psi(\xi)}\hat P^T(\xi,\xi^*)\ket{\psi(\xi^*)}<0.
\label{equ:expect}
\end{equation}
Denoting the pure state on the Bloch sphere which has the smallest overlap with the 
transpose of the initial state of the qubit as $(u,v):=(-\sqrt{(1-z/r)/2},e^{-i\phi}\sqrt{(1+z/r)/2})$, 
it turns out that
\begin{equation}
 \ket{\psi(t;\xi^*)}=
\begin{pmatrix}
 e^{-\left(\xi|a+b\right)/2+i\Omega t/2}\,u\\-e^{\left(\xi|a+b\right)/2-i\Omega t/2}\,v
\end{pmatrix}
\label{equ:test-state}
\end{equation}
fulfills this task. This choice of state is further elaborated upon in section~\ref{sec:sme}.
By solving the integral~(\ref{equ:expect}) one finds the following expression for the expectation value:
\begin{equation}
\begin{split}
\mathcal{E}^\PT(t)=\frac{1}{2\left(\n+1\right)}&\left(e^{-A(t)+\frac{\overline{S}(t)}{2}}\left[1
-\frac{z^2}{r}\right]\right.\\
&\qquad-\left.e^{B(t)-\frac{\overline{S}(t)}{2}}\frac{(x^2+y^2)}{r}\right),
\end{split}
\label{equ:negative-expect}
\end{equation}
where $\overline{S}$ arises from $S=A+B$ by inverting the coefficients
$(\n_\lambda+1)/\n_\lambda\rightarrow\n_\lambda/(\n_\lambda+1)$. This quantity becomes negative,
i.e. the total state is entangled, if the following condition is fulfilled:
\begin{equation}
  E(T,t):=S(T,t)-\overline{S}(T,t)>\ln\left[\frac{r-z^2}{x^2+y^2}\right].
\label{equ:condition-ent}
\end{equation}
Note that $\left(r-z^2\right)/(x^2+y^2)\ge 1$ and hence the quantity on the
r.h.s.~(\ref{equ:condition-ent}) is always non-negative.

As before with inequality~(\ref{equ:condition-sep}) for separability, for any initial state, this new
inequality~(\ref{equ:condition-ent}) defines an area in the temperature-time plane in which entanglement
of the total state is assured. As before,
this is a sufficient but not a necessary condition for entanglement.

In terms of the spectral density the relevant function reads
\begin{equation}
\begin{split}
E(T,t)=&\;8\int_0^t\.ds\int_0^s\.d\tau\int_0^\infty\.\.d\omega\\
&\quad\times J(\omega)\sinh\left[\hbar\omega/kT\right]\cos[\omega(s-\tau)].
\end{split}
\label{equ:E_t_continuum}
\end{equation}
In the limit of high temperatures and large times, $E(T,t)$ becomes independent of time,
$E(T,t)\rightarrow E(T)\approx 8\kappa\hbar\omega_c/k_B T$. As a consequence, for every initial state there is a
temperature above which our criterion is no longer able to detect entanglement.
As we will show in figures~\ref{fig:sep-ent-dec} (a) and (b) for low temperatures and relatively short
times the criteria for separability and entanglement cover almost the entire $(T,t)$-plane and only little
``unknown territory'' (the white areas) remains.

\subsection*{Separability and Entanglement in the Strong Coupling Regime}
\begin{figure}
\includegraphics[width=0.47\textwidth]{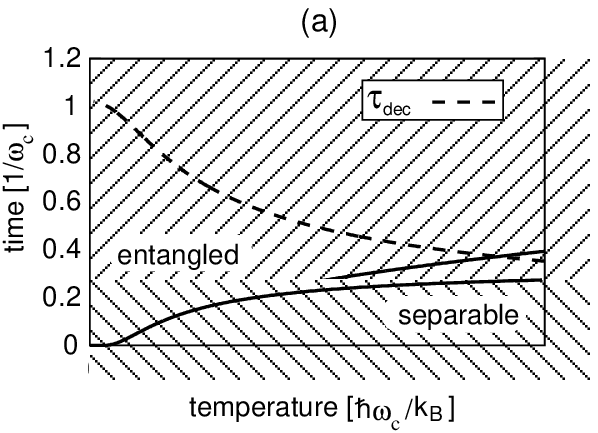}\\
\includegraphics[width=0.47\textwidth]{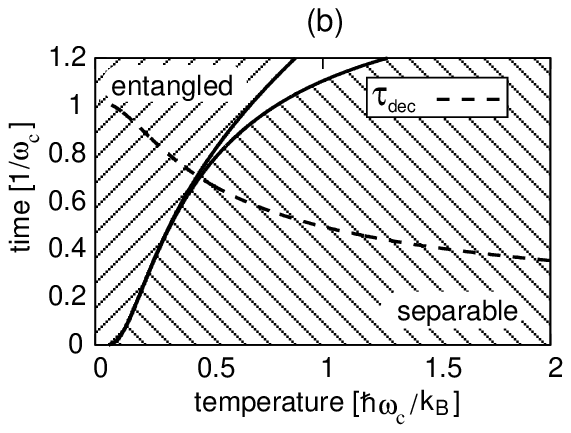}
 \caption{System-environment correlations vs decoherence time - strong interaction regime:
(a) For initial states of high purity (here $r=0.9$) decoherence
mainly takes place after the total state has become entangled.
(b) for mixed initial conditions decoherence can occur in both domains,
here shown for an initial state with purity $r=0.1$.}
 \label{fig:sep-ent-dec}
\end{figure}

In fig.~\ref{fig:sep-ent-dec}
we relate the decoherence time $\tau_\dec(T)$ to our criteria for separability and
entanglement. To ensure that entanglement builds up fast, we here resort to
the \emph{strong coupling} regime ($\kappa=1$). It
becomes apparent that for an initial qubit state that is pretty pure
($r=0.9$), in the considered low temperature range, decoherence is complete only after entanglement has built up (a).
By contrast, for an initial qubit mixed state with purity 
$r=0.1$ (depending on temperature) decoherence can occur \emph{before} or \emph{after}
the two subsystems were able to entangle (b).

These results show that in general there is no
direct connection between decoherence and the formation of system-reservoir 
entanglement: surely, if the qubit is prepared in a pure 
state, decoherence is always accompanied by entanglement 
of the total state, independently of temperature. If the initial state is 
already a completely incoherent mixture, the two subsystems will never get entangled.
In all other cases there are two disjoint domains of temperature
implying decoherence {\it before} and {\it after} the qubit was able
to entangle with its environment, respectively.

\begin{figure}
\includegraphics[width=0.47\textwidth]{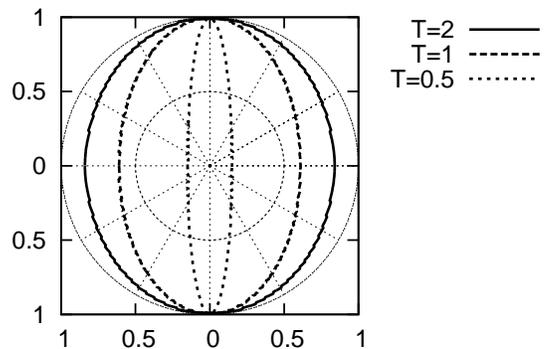}
\caption{Separable decoherence:
cut through the Bloch sphere at $\phi=0$. All states \emph{inside} the area enclosed by a curve of given temperature
decohere while the total state is still separable.
Temperatures in units of $\hbar\omega_c/k_B$.}
 \label{fig:scan_blochsphere_sep}
\end{figure}

\begin{figure}
\includegraphics[width=0.47\textwidth]{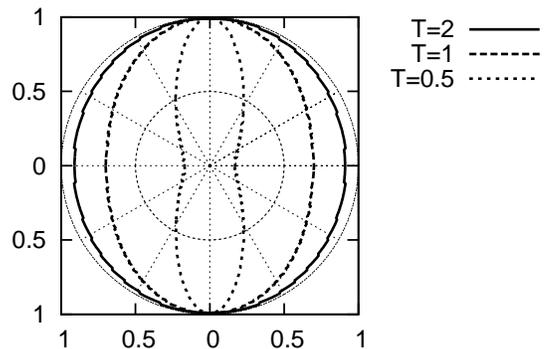}
\caption{Entangling decoherence:
cut through the Bloch sphere at $\phi=0$. All states \emph{outside} the area determined by a curve of given temperature
decohere after entanglement has built up.
Temperatures in units of $\hbar\omega_c/k_B$.}
 \label{fig:scan_blochsphere_ent}
\end{figure}

We now want to further elucidate the nature of decoherence depending on the initial state of the qubit.
In the cut through the Bloch sphere of fig.~\ref{fig:scan_blochsphere_sep}
all states \emph{inside} the area enclosed by a curve of given temperature
decohere while the total state is still separable.
In similar vein, in fig.~\ref{fig:scan_blochsphere_ent}
all states \emph{outside} the area determined by a curve of given temperature
decohere after entanglement has built up.

As the coupling to the environment distinguishes the south and north pole of the Bloch sphere as robust states,
for a given purity $r$, 
there is no global value of temperature $T$ defining the cut between 
``separable'' and ``entangling'' decoherence. Rather, 
states close to the poles tend to decohere while the total state is still separable.
States near the equator, by contrast, tend to decohere involving entanglement.

\subsection*{Oscillations between Separability and Entanglement}
\begin{figure}
\includegraphics[width=0.47\textwidth]{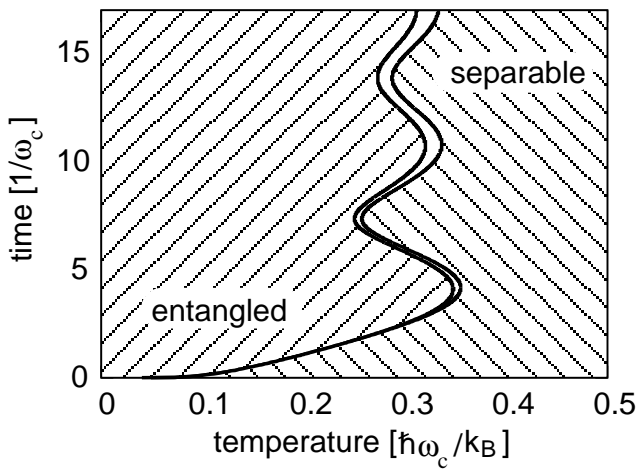}
\caption{Oscillatory behavior in correlation dynamics: at low
temperatures one can find conditions where separable and entangled domains
alternate, here shown in the \emph{weak interaction} regime
($\kappa=10^{-3}$) for a qubit initial state 
with purity $r=0.95$ ($z=\phi=0$).}
\label{fig:oscillations}
\end{figure}
As already mentioned in the beginning,
the dynamics of system-environment correlations can be
quite rich. In the low temperature regime, in particular, we are able to detect
conditions where the total state alternates between being entangled and
separable.
In fig.~\ref{fig:oscillations} we show the correlation dynamics for
an initial qubit state with purity $r=0.95$. It becomes apparent that for
temperatures around $T=0.3\,\hbar\omega_c/k_B$ the nature of the correlations
changes oscillatorily between classical and
quantum as a function of time. Although the special form of these oscillations depends on the form of
the spectral density $J(\omega)$, we found numerically that similar behavior exists for different
choices of $J$.

\section{Quality of Entanglement Detection for a Single Mode ``Environment''}
\label{sec:sme}
\begin{figure}
\includegraphics[width=0.47\textwidth]{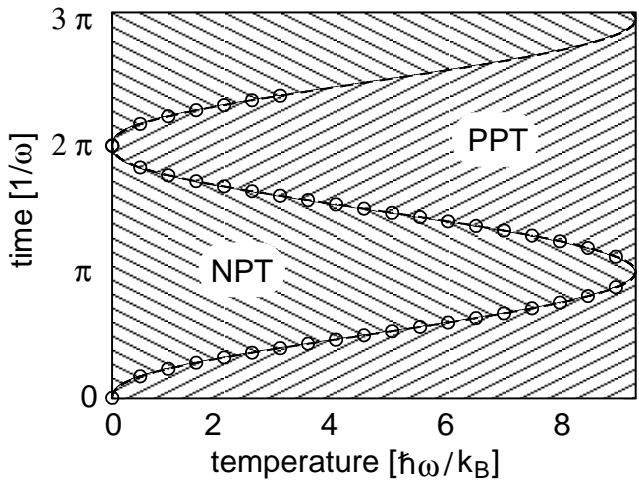}
\caption{Single mode environment - comparison to numerics:
temperature dependence of the exact instant $\tau_\crit(T)$ at
which the total state changes from PPT to NPT, and vice versa (points). The result obtained
by our criterion~(\ref{equ:condition-ent}) is displayed by the shaded areas. Obviously, there is perfect agreement.
Qubit initial state with purity $r=0.75$, $z$-coordinate $z=0.2$ and azimuth angle $\phi=0$.}
\label{fig:numerics}
\end{figure}
\vspace{4em}
Fig.~\ref{fig:oscillations} demonstrates that -- at low temperatures at least -- our
separability~(\ref{equ:condition-sep}) and entanglement~(\ref{equ:condition-ent}) criteria are quite good:
the ``unknown territory'' of correlations is very small.
To also estimate its accuracy for higher temperatures we study
entanglement dynamics of the total state in the case of a
single mode "environment", where the full dynamics can be accessed numerically.
We are interested in the exact instant $\tau_\crit(T)$ at which the definiteness of the partial
transpose $\rho_\tot^\PT$ ceases to be positive (PPT) and negative eigenvalues emerge (NPT).
On the other hand
eq.~(\ref{equ:condition-ent}) allows one to analytically derive the instant
$\tau_\ent(T)$ from which on we are able to detect entanglement with our test state determined from
eq.~(\ref{equ:test-state}).

In fig.~\ref{fig:numerics} we compare the temperature dependence of these characteristic
times for an initial state with purity $r=0.75$.
It turns out that there is an excellent agreement between the exact numerical
and the analytical result, even at high temperatures.
We are confident that also in the case of larger environments, our entanglement criterion~(\ref{equ:condition-ent})
is able to very efficiently detect a negative partial transpose.
Clearly, it is an interesting and open question, whether the white areas in our $(T,t)$-diagrams correspond to
entangled, bound entangled or separable total states.

\section{Conclusions}
We have investigated the dynamics and nature of system-environment correlations of a 
decohering qubit with arbitrary (mixed) initial state coupled to a bath of quantum harmonic oscillators. 
When the reduced
dynamics can be described by a Markov master equation of Lindblad type, we 
found that for almost all qubit initial states inside the Bloch sphere decoherence is complete
long before the build-up of any system-reservoir entanglement: the total state is still separable.

By contrast, in the strong coupling regime the correlation dynamics was found to be more complex: dependent
on initial state and temperature of the surrounding heat bath, decoherence can occur before or
after the qubit is able to entangle with its environment.
At very low temperatures we were even able to detect conditions where the total state alternates between
separable and entangled domains as a function of time.

Finally, we have compared our criterion for system-reservoir entanglement with
exact numerical results for the occurrence of a negative partial transpose
in the case of a single mode environment and found excellent agreement.
We therefore believe to have found a reliable criterion for system-bath entanglement.
Clearly, it is desirable to further investigate the nature of 
system-environment correlations in the white regions in order to cover
the whole range of temperatures and times. Moreover, we
are confident that our approach may be applied to more general
open quantum system dynamics.

\section*{Acknowledgements}
The authors thank Julius Helm, G\"unter Plunien, Lajos Di\'osi, Patrick Navez,
and Ting Yu for fruitful discussions and hints.
A.P. acknowledges support from the International Max Planck Research School 
(IMPRS) Dresden and from GSI Darmstadt.

\end{document}